\documentclass[a4paper,twoside]{article}
\usepackage{graphicx,psfig,epsfig}
\usepackage{amsfonts}
\usepackage{amssymb}
\usepackage{amsmath}
\usepackage{CoLe}

\newcommand{\CoLetitle}%
{Structural and Electronic Properties of Gold Clusters}

\newcommand{\CoLeauthor}
{\textbf{Denitsa Alamanova,\footnote{e-mail: deni@springborg.pc.uni-sb.de} 
Yi Dong,\footnote{e-mail: y.dong@mx.uni-saarland.de} 
Habib ur Rehman,\footnote{e-mail: haur001@rz.uni-saarland.de}\\  
Michael Springborg,\footnote{Corresponding author. e-mail: m.springborg@mx.uni-saarland.de}
and Valeri G.\ Grigoryan\footnote{e-mail: vg.grigoryan@mx.uni-saarland.de} }
\\[1ex]
Physical and Theoretical Chemistry,\\ University of Saarland,\\
 66123 Saarbr\"ucken,\\ Germany}

\newcommand{\yearnumber}{2005}
\newcommand{\volumenumber}{1}
\newcommand{\issuenumber}{1}
\newcommand{\firstpage}{1}
\newcommand{\lastpage}{12}
\begin{document}

\maketitle
\markboth{\CoLeauthorshort{D.Alamanova, Y.Dong, H.u.Rehman, M.Springborg, V.G.Grigoryan}}%
    {\CoLetitleshort{Au Clusters}}

\centerline{Received 30 June, 2005; accepted in revised form 30
June, 2005}

\begin{abstract}
We study the structure and energetics of Au$_N$ clusters by means of 
parameter-free density-functional calculations ($N\le 8$), jellium calculations
($N\le 60$), embedded-atom calculations ($N\le 150$), and parameterized 
density-functional calculations ($N\le 40$) in combination with different 
methods for determining the structure of the lowest total energy. 
By comparing the results from the different approaches, effects due to 
geometric packing and those due to the electronic orbitals can be identified.
Different descriptors that highlight the results of the analysis are presented
and used.
\end{abstract}

\begin{keywords}
Gold clusters, structure, stability, density-functional calculations, embedded-atom calculations
\end{keywords}

\begin{PACS}
36.40.-c, 36.90.+f, 61.46.+w, 73.22.-f
\end{PACS}

\section{Introduction}

Clusters of gold atoms have become the maybe mostly studied class of 
clusters \cite{pekka}, partly due to the possibility to apply them in electronic devices \cite{md95}, 
nanomaterials \cite{rlw99}  and catalysis \cite{as99}. Despite this popularity only
little consensus has been reached concerning the structure of these clusters. Many 
studies devoted to this issue use combinations of experimental and theoretical methods 
\cite{ha03,sg02,mn05}.

From a theoretical point of view, gold clusters offer an additional challenge due
to the importance of relativistic effects, most notably the strong spin-orbit couplings. 
Thus, in parameter-free, electronic-structure calculations, one has to use
special, relativistic potentials. With those, the smallest gold clusters are found 
to be planar \cite{ha03,br99,gr00,ha00,wa02,avw05,fr05}, whereas their exclusion 
leads to three-dimensional structures \cite{do98,wi00}.

However, for not-too-small gold clusters, it becomes increasingly difficult to apply
parameter-free, electronic-structure methods in the calculation of the properties of
the gold clusters, partly because the computational demands scale with the size of the
system to at least the third power, and partly because the number of metastable structures
grows very rapidly with cluster size. Thus, in parameter-free studies one often has to
make significant assumptions on the structure of the system, as for instance is the 
case in the study of H\"aberlen {\it et al.} \cite{ha97}. 
One of the greatest disadvantages of the first-principles
methods is their incapability of optimizing large number of randomly generated
initial structures and thereby determining the true global total-energy minimum. One example
of this is provided by Au$_7$ and Au$_8$ for which the structure was predicted in 2000 by H\"akkinen
and Landman \cite{ha00} to be a planar structure with D$_{2h}$
symmetry and a three-dimensional capped tetrahedron, respectively. Three years later,
H\"akkinen {\it et al.} \cite{ha03} showed that the lowest-total-energy structure of
Au$_7^-$ corresponds to a planar structure
consisting of a rhombus, capped with an additional atom on three of its sides,
and that the ground state of Au$_8^-$ is the same rhombus, with its 4 sides
capped. In 2005, the results for Au$_8$ were confirmed by Walker \cite{avw05}
and Remacle {\it et al.} \cite{fr05}, whereas both works found a planar capped
hexagon to be the global minimum of neutral Au$_7$.

Approximate methods may provide a useful alternative to the parameter-free methods. They
are computationally less demanding, thus allowing for a detailed search in structure space
so that structures for clusters with well above 100 atoms can be predicted in an unbiased
way. On the other hand, being approximate it is not obvious how reliable they are. It is
the purpose of this work to address this issue. To this end, gold clusters provide an excellent
playground, partly because of the large uncertainty concerning their structure in combination
with the large amount of studies on these clusters, but also partly because for clusters both 
geometric packing effects and electronic shell effects may be responsible for the occurrence
of certain particularly stable clusters (the so-called magic numbers). The approximate methods
often make different approximations on the relative importance of these two effects.

One class of approximate methods is formed by the embedded-atom methods (EAM) that only
indirectly includes electronic effects and, therefore, first of all (but not exclusively)
put emphasis on packing effects. Both the EAM \cite{clc97,rb99,cc97}, the 
Sutton-Chen \cite{do98}, the Murrell-Mottram \cite{wi00}, and the many-body 
Gupta potential \cite{ilg96,ilg98,ilg99,km99,tl00,sd02} (that all share the property of 
including electronic effects only very approximately) have all been applied in 
unbiased structure optimizations for gold clusters with up to 80 atoms. 

One of the, maybe, surprising outcomes of these studies is that the results depend very
sensitively on the applied method, i.e., on the (more or less) approximate description of
the interatomic interactions and on the method for structure optimization, see, e.g.,
\cite{bf05,nanjing}. One reason may be a subtle interplay between electronic and geometric
effects, i.e., that the particularly stable structures of Au$_N$ clusters are dictated
partly by the closing of electronic shells and partly by geometric packing effects. Here,
the various potentials give different relative importance to the two effects.

The purpose of this contribution is to discuss general methods for calculating the 
properties of clusters, using gold clusters as the prototype. In parallel we shall also
discuss the special properties of the gold clusters, specifically, with special 
emphasis on the issue above, i.e., whether electronic or packing effects are important
in dictating the particularly stable clusters, and, moreover, how the different 
more or less accurate methods perform in calculating the properties of the clusters.

\section{Methods}

\subsection{Total-energy methods}

The smallest gold clusters with up to eight atoms were treated with parameter-free
electronic-structure calculations using the \textsc{Gaussian03} program package \cite{gauss}. We 
performed density-functional (DFT) calculations using the generalized-gradient approximation (GGA)
of Perdew, Burke, and Ernzerhof \cite{pbe,jpp97}. These calculations treat, in principle, 
all types of interactions, i.e., electronic and geometric effects, at an exact level.

In addition we also performed self-consistent, electronic-structure calculations on spherical
clusters where only the 11 ($5d$ and $6s$) valence electrons per Au atom were treated
explicitly, whereas all core electrons and the nuclei were smeared out to a uniform 
jellium background \cite{jel1,jel2}. This model focuses essentially only on electronic
effects. 

As an alternative we also considered the embedded-atom method (EAM) in the parameterization
of Voter and Chen \cite{vo87,vo93,vo95}. According to this method, the total energy for a
system of $N$ atoms is written as
\begin{eqnarray}
E_{\rm tot}&=&\sum_{i}E_i\nonumber\\
E_i&=&F_i(\rho_i^h)+{\frac{1}{2}\sum_{j(\ne i)}\phi_{ij}(r_{ij})}\nonumber\\
\rho_i^h&=&\sum_{j \,(\ne i)} \rho_i^a (r_{ij}),
\end{eqnarray}
i.e., as a sum of atomic components, each being the sum of two terms. The first term is 
the energy that it costs to bring the atom of interest into the electron density provided
by all other atoms, and the second term is a pair-potential term.
Both terms are assumed depending only on the distances between the neighbouring atoms, and
do therefore not include any directional dependence. Accordingly, the EAM emphasizes geometrical
effects, whereas electronic effects are included only very indirectly.

Furthermore, we used the density-functional tight-binding method (DFTB) as developed by 
Seifert and coworkers \cite{gs92,gs96}. With this method, the binding energy is written as
the difference in the orbital energies of the compound minus  those of the isolated atoms, i.e., as 
\begin{equation}
\sum_i\epsilon_i-\sum_m\sum_i\epsilon_{mi}
\end{equation} 
(with $m$ being   an atom index and $i$ an orbital index), augmented with pair potentials, 
\begin{equation}
\sum_{m_1\ne m_2}
  U_{m_1,m_2}(\vert\vec R_{m_1}-\vec R_{m_2}\vert)
\end{equation}
(with $\vec R_m$ being the position of the $m$th atom). 
In calculating the orbital energies we need the Hamilton matrix elements
  $\langle\chi_{m_1n_1}\vert\hat H\vert\chi_{m_2n_2}\rangle$ and the overlap matrix elements
  $\langle\chi_{m_1n_1}\vert\chi_{m_2n_2}\rangle$. Here, $\chi_{mn}$ is the $n$th atomic
  orbital of the $m$th atom. The Hamilton operator contains the kinetic-energy operator as
  well as the potential. The latter is approximated as a superposition of the potentials of
  the isolated atoms, 
\begin{equation}
V(\vec r)=\sum_m V_m(\vert\vec r-\vec R_m\vert),
\end{equation}
 and subsequently
  we assume that the matrix element $\langle\chi_{m_1n_1}\vert V_m\vert\chi_{m_2n_2}\rangle$
  vanishes unless at least one of the atoms $m_1$ and $m_2$ equals $m$. Finally, the pair 
potentials $U_{m_1,m_2}$ are obtained by
  requiring that the total-energy curves from parameter-free density-functional calculations
  on the diatomics are accurately reproduced.

\subsection{Structure determinations}

We used several different methods in determining the structures of the clusters. In the 
\textsc{Gaussian03} calculations the clusters were so small that it was possible to 
determine the structures of the lowest total energy simply through searching in the 
structure space. In the jellium calculations there is per construction no structure and 
the system has a spherical symmetry. 

In the EAM calculations we optimized the structure using our own {\it Aufbau/Abbau} method
\cite{aa1,aa2,aa3}. The method is based on simulating experimental conditions, where clusters
grow by adding atom by atom to a core. By repeating this process {\bf very} many times and 
in parallel also removing atoms from larger clusters, we can identify the structures of 
the lowest total energy. 

Finally, in the DFTB calculations we used two different approaches. In one approach the 
structures of the EAM calculations were used as input for a local relaxation, i.e., only
the nearest local-total-energy minimum was identified. In another set of calculations, 
we optimized the structures using the so-called genetic algorithms \cite{ga1,ga2,ga3}.
Here, from a set of structures we generate new ones through 
cutting and pasting the original ones. Out of the total set of old and new clusters
those with the lowest total energies are kept, and this process is repeated until the
lowest total energy is unchanged for a large number of generations.

\section{Results}

The smallest possible cluster is the Au$_2$ molecule. For this we show in Table \ref{I}
the calculated bond length and binding energy from the different methods in comparison 
with experimental values. Notice that the DFTB method has been parameterized to reproduce
results from parameter-free density-functional calculations on precisely the dimer and is,
therefore, for the dimer accurate.

\begin{table}[hc]
\caption{A comparison between the experimental and the calculated bond length
and binding energy of the Au dimer obtained with the parameter-free density-functional
calculations and the EAM method.}
\label{I}
\begin{tabular}{lccccc}
\hline\noalign{\smallskip}

Au$_2$ && R$_e$, \AA && E$_b$, eV\\
\noalign{\smallskip}\hline\noalign{\smallskip}

DFT && 2.55 && 2.22\\
EAM && 2.40 && 2.29\\
{\it EXP} && 2.47 && 2.29\\

\noalign{\smallskip}\hline
\end{tabular}
\end{table}

\unitlength1cm
\begin{figure}[tbp]
\begin{picture}(18,9)
\put(0,6){\psfig{file=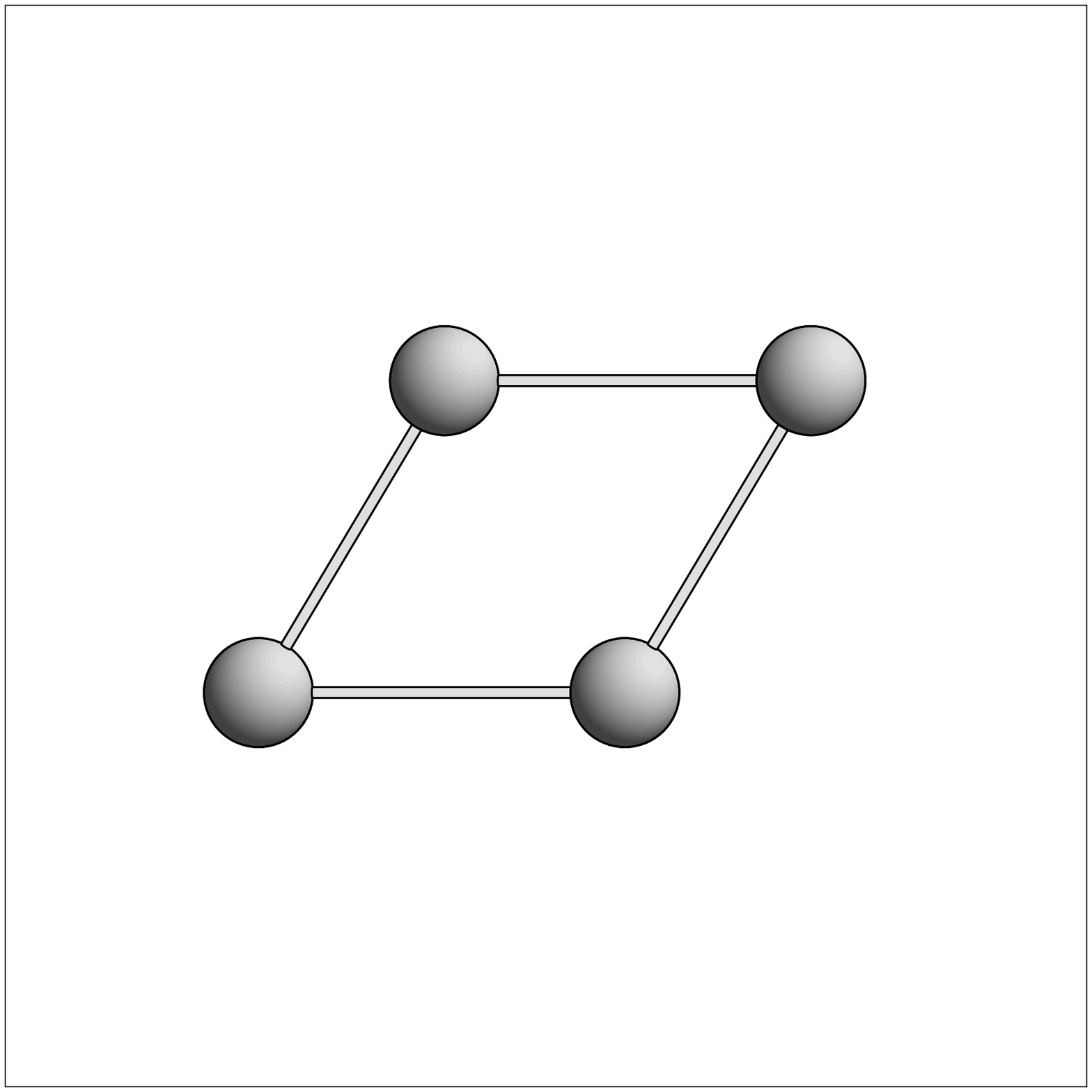,width=2.5cm}}
\put(3,6){\psfig{file=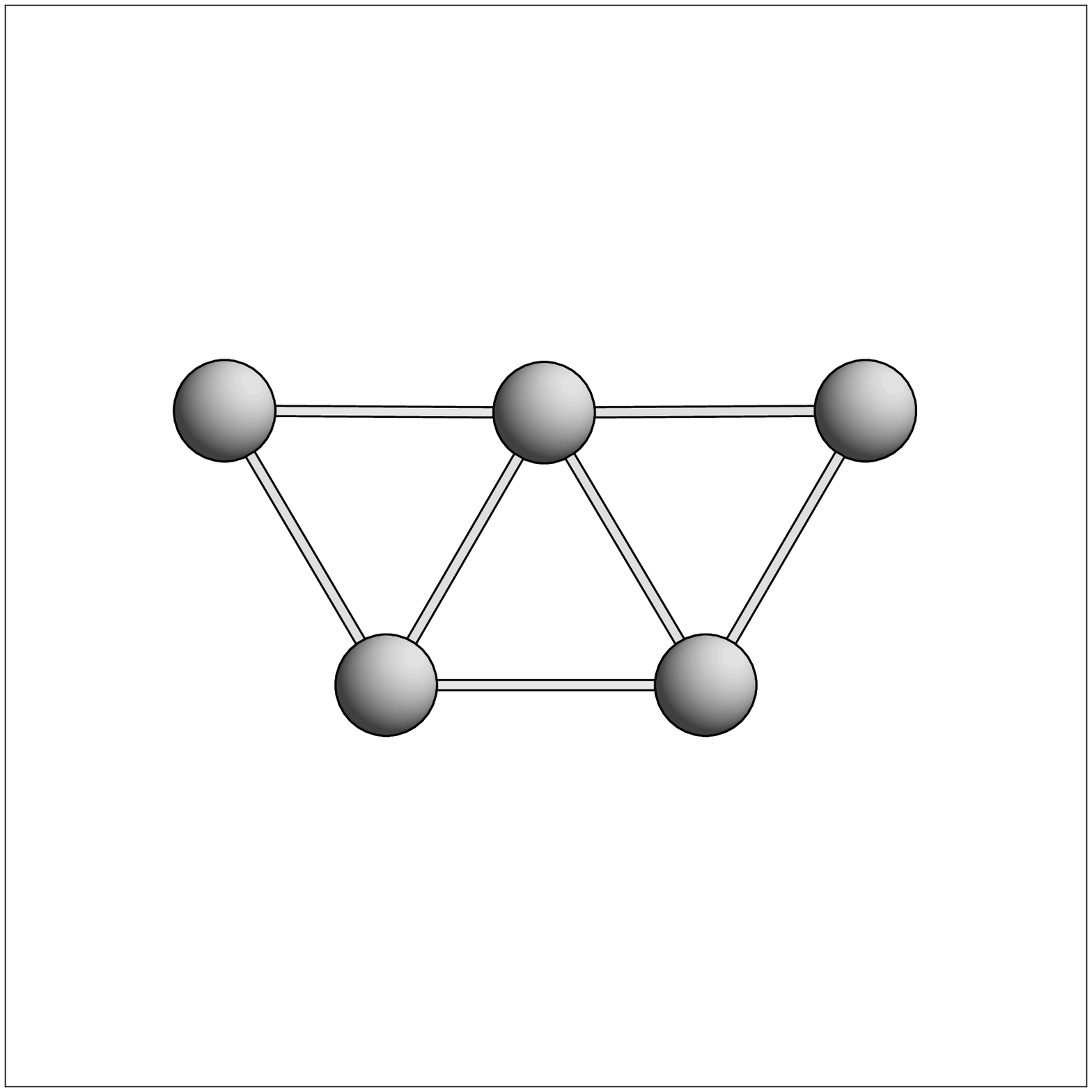,width=2.5cm}}
\put(6,6){\psfig{file=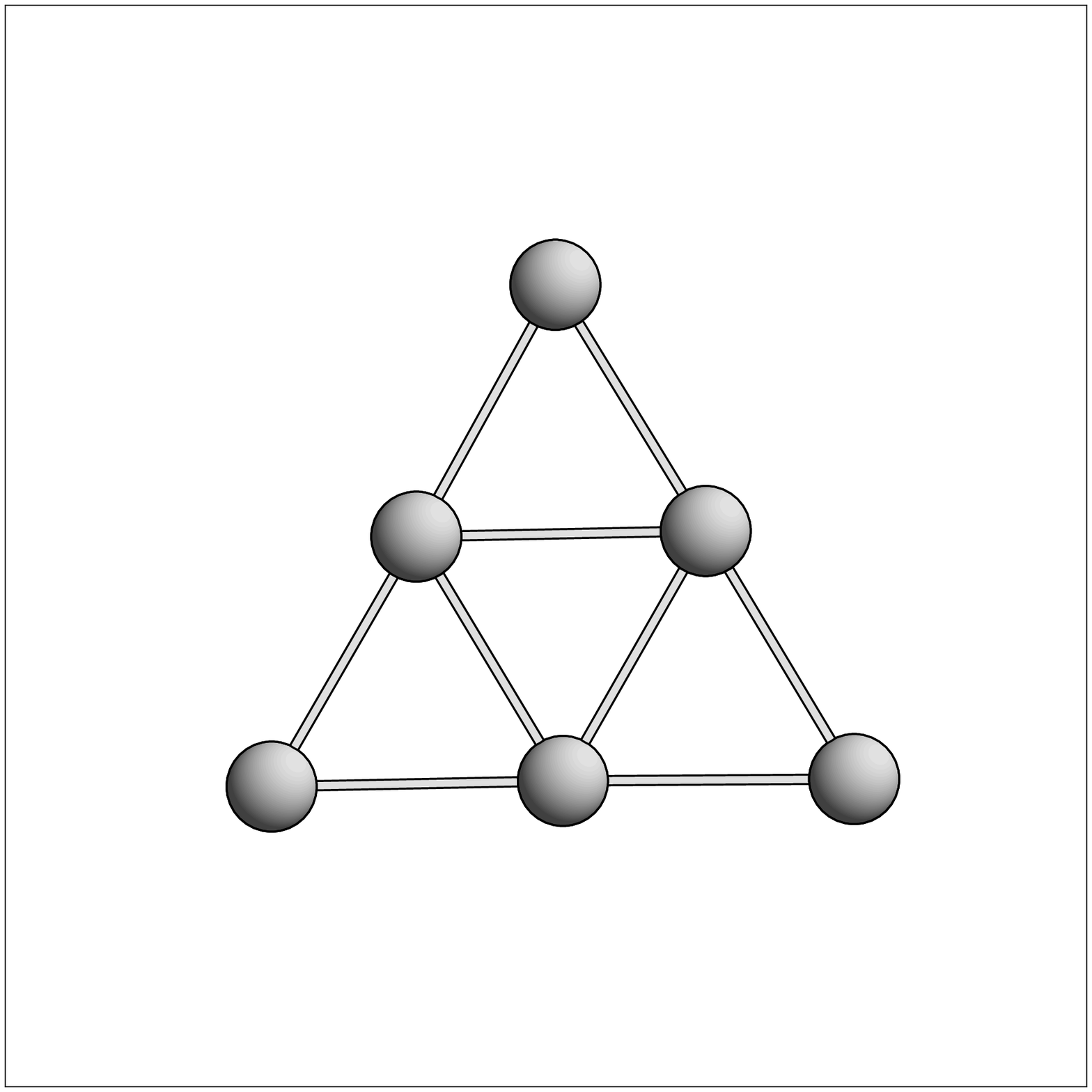,width=2.5cm}}
\put(9,6){\psfig{file=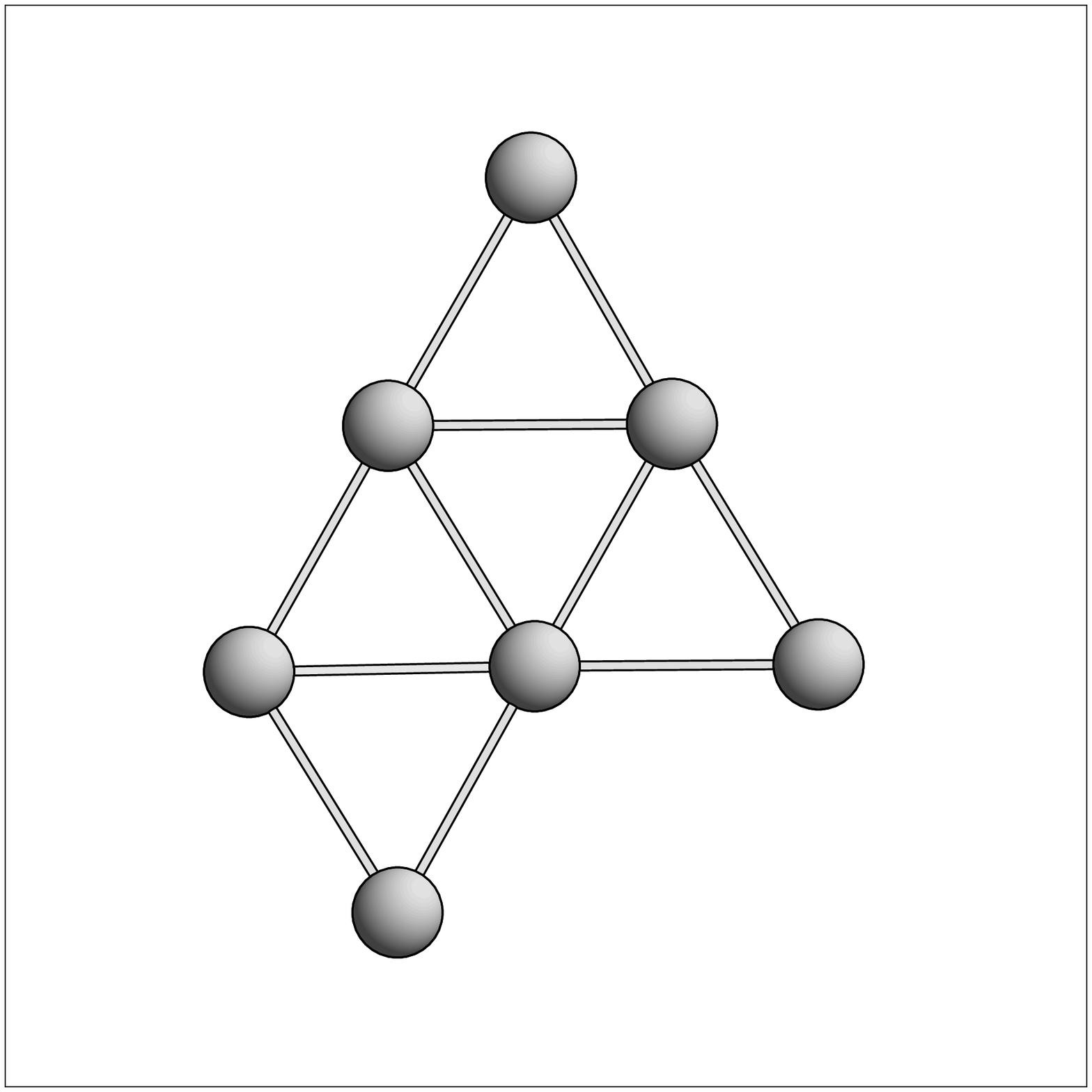,width=2.5cm}}
\put(12,6){\psfig{file=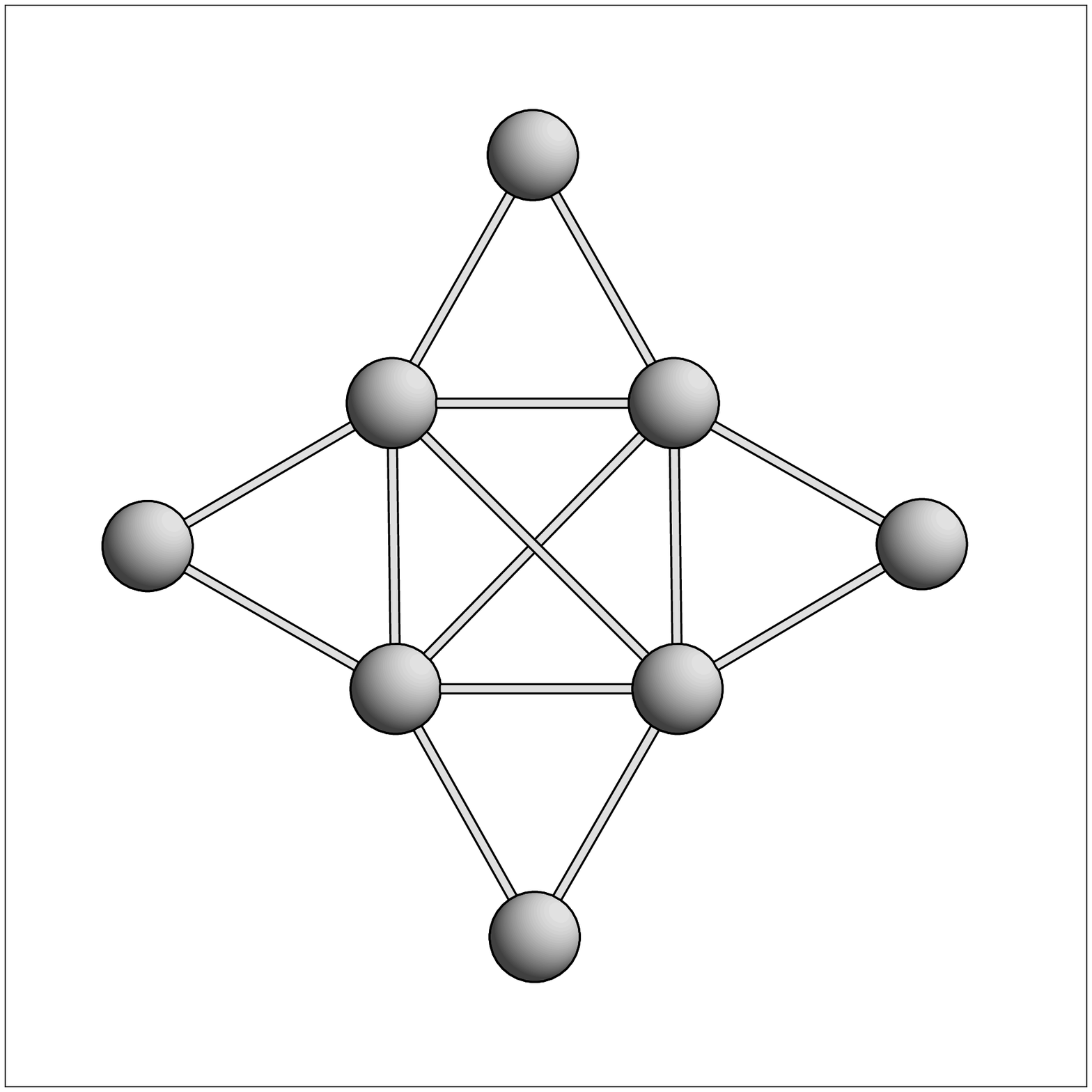,width=2.5cm}}
\put(0,3){\psfig{file=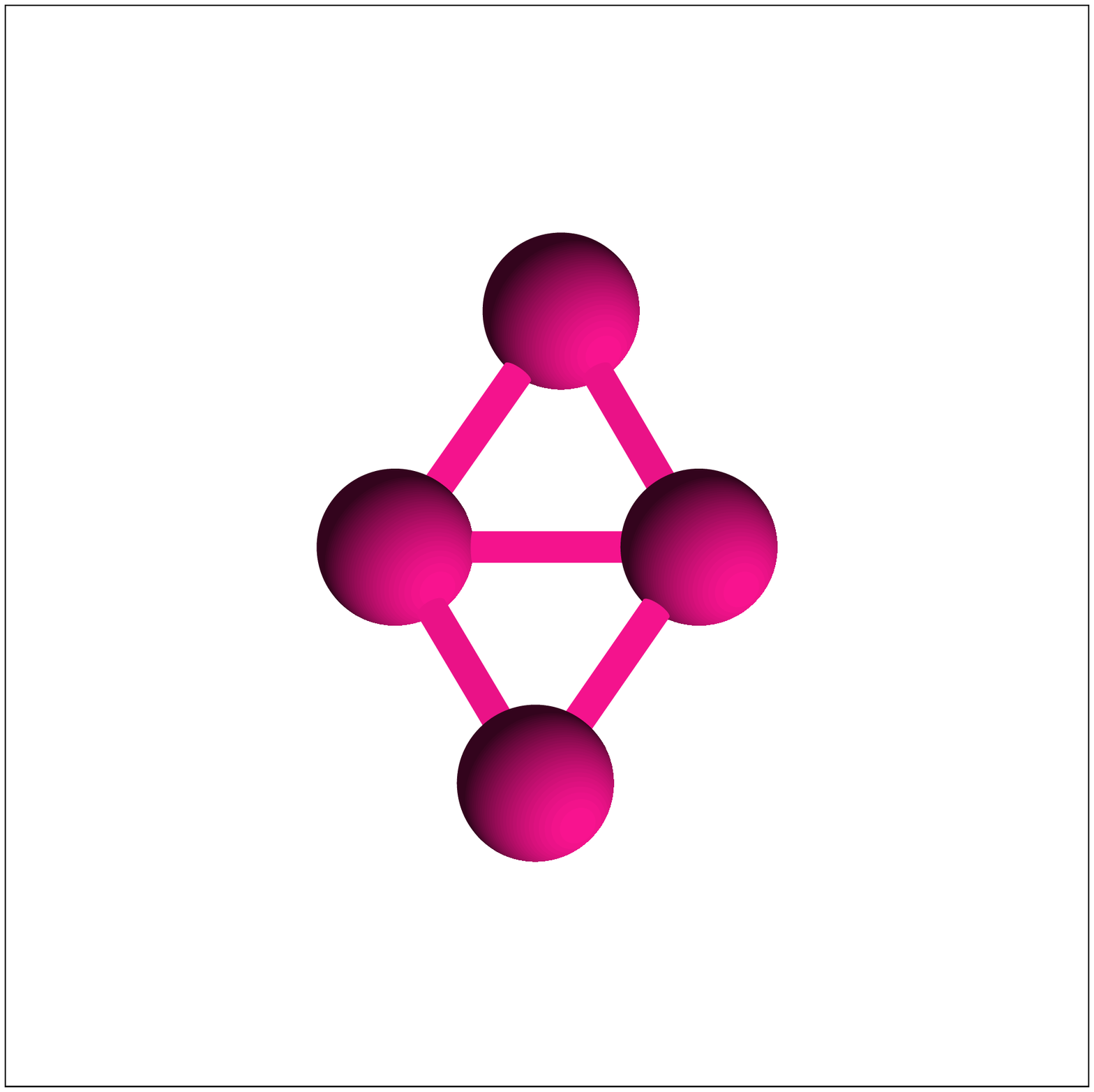,width=2.5cm}}
\put(3,3){\psfig{file=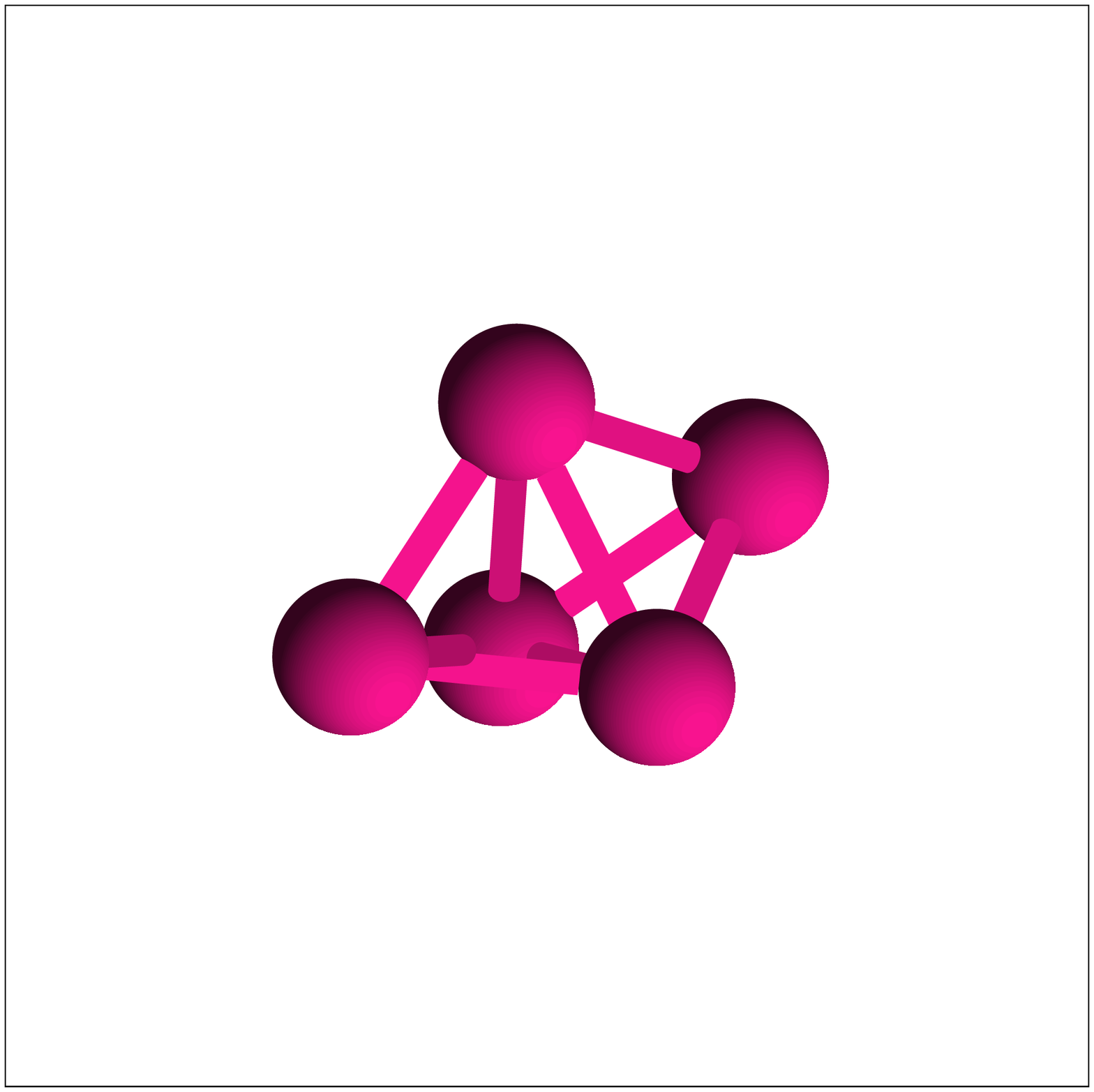,width=2.5cm}}
\put(6,3){\psfig{file=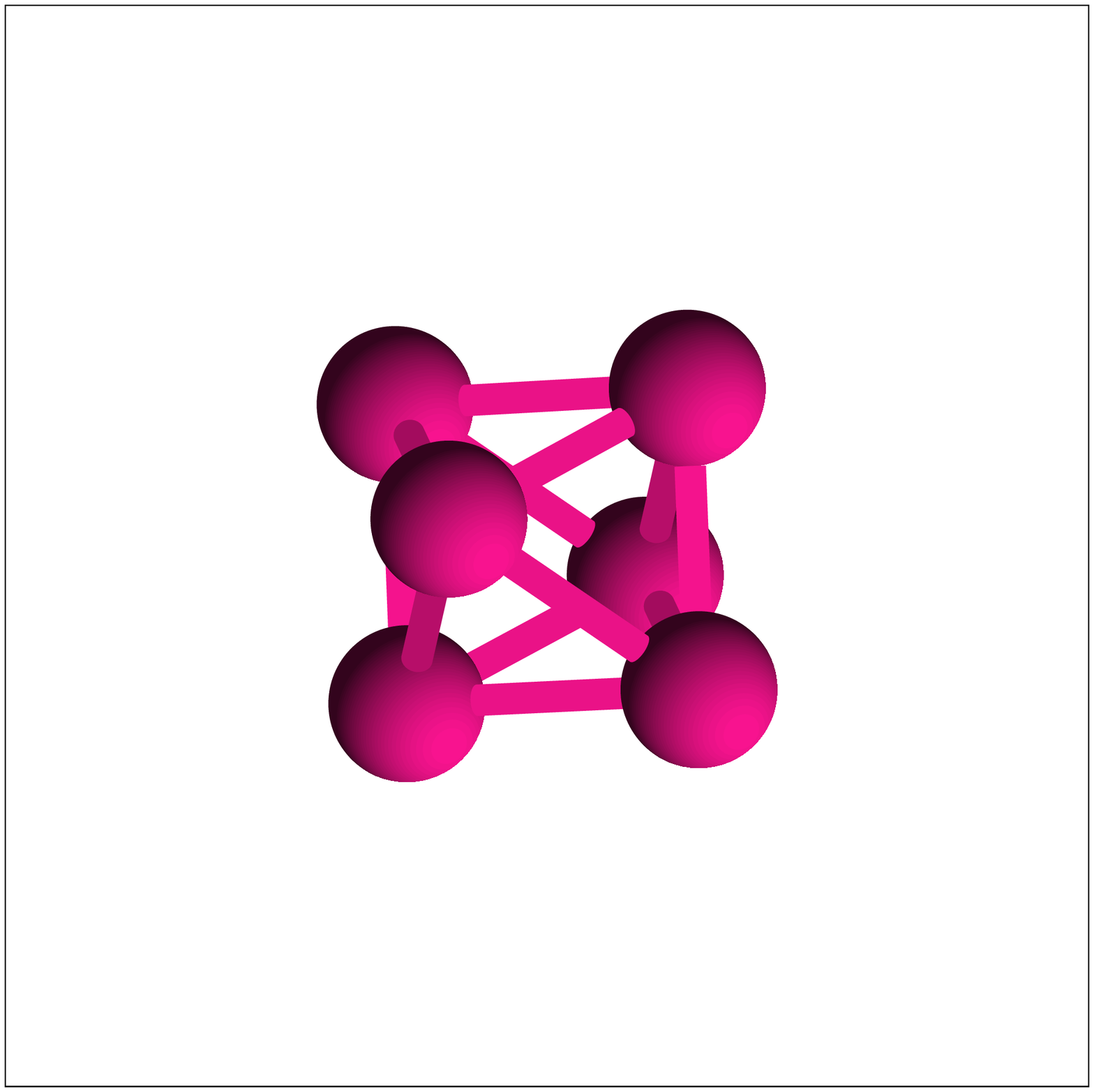,width=2.5cm}}
\put(9,3){\psfig{file=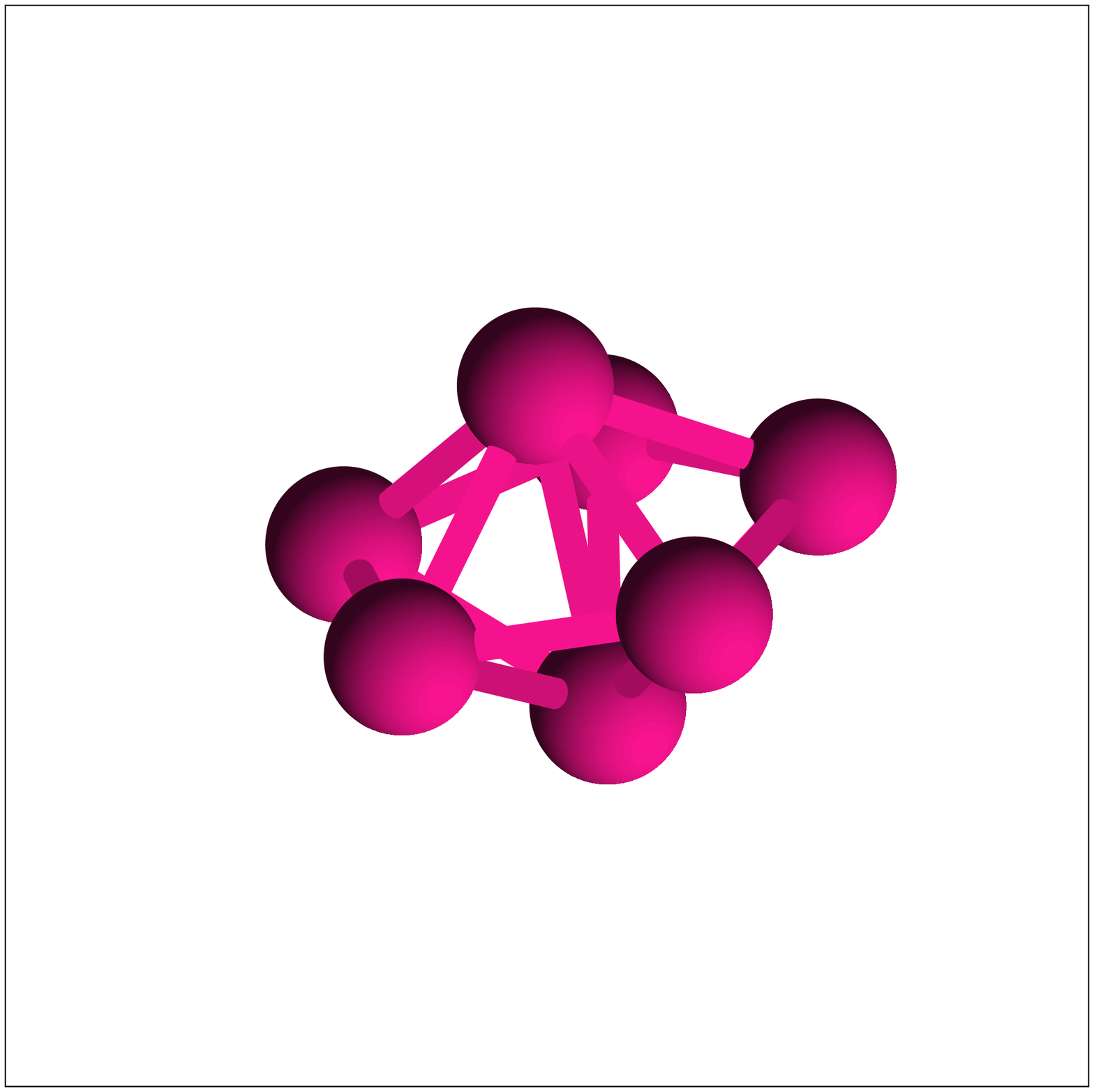,width=2.5cm}}
\put(12,3){\psfig{file=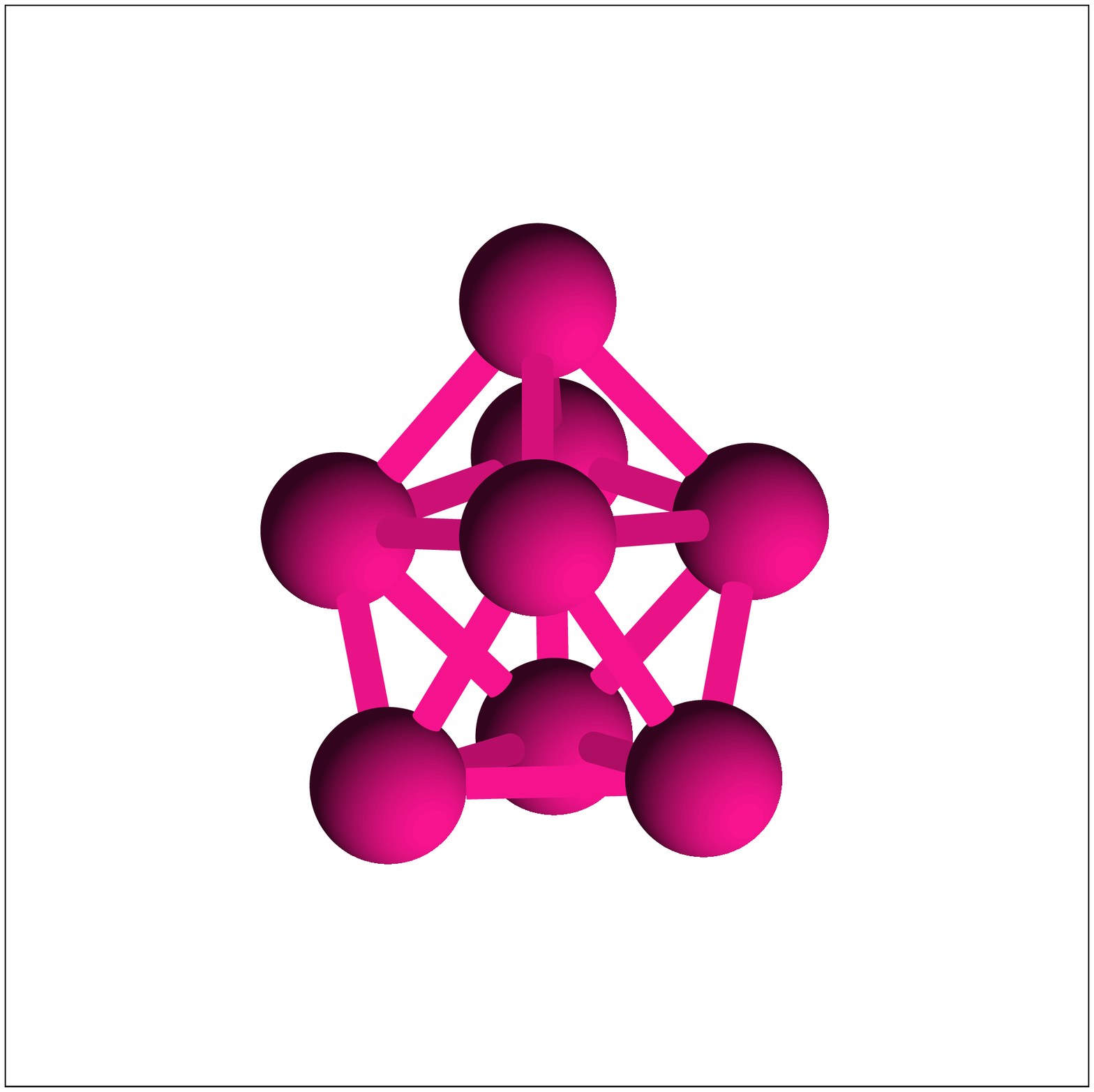,width=2.5cm}}
\put(0,0){\psfig{file=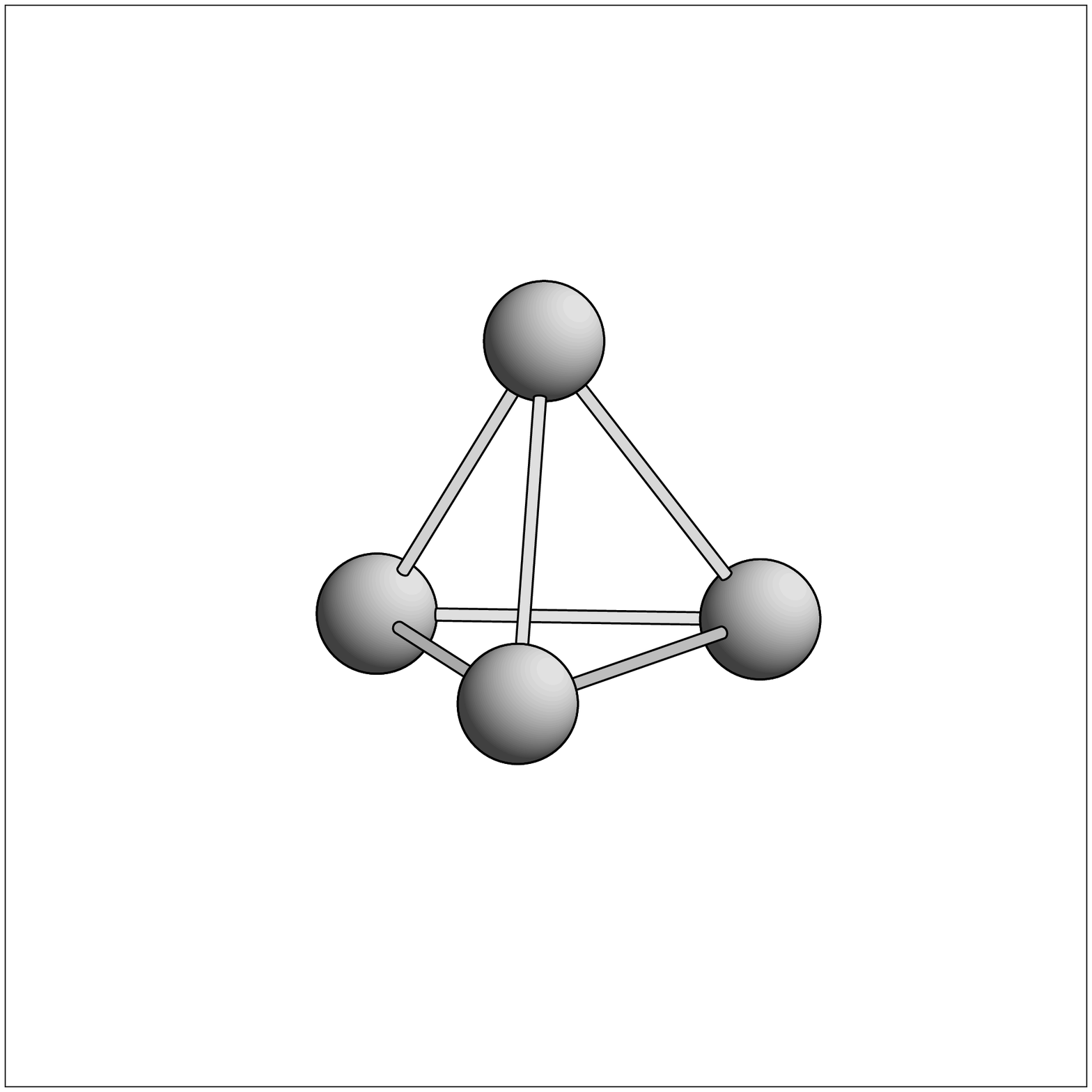,width=2.5cm}}
\put(3,0){\psfig{file=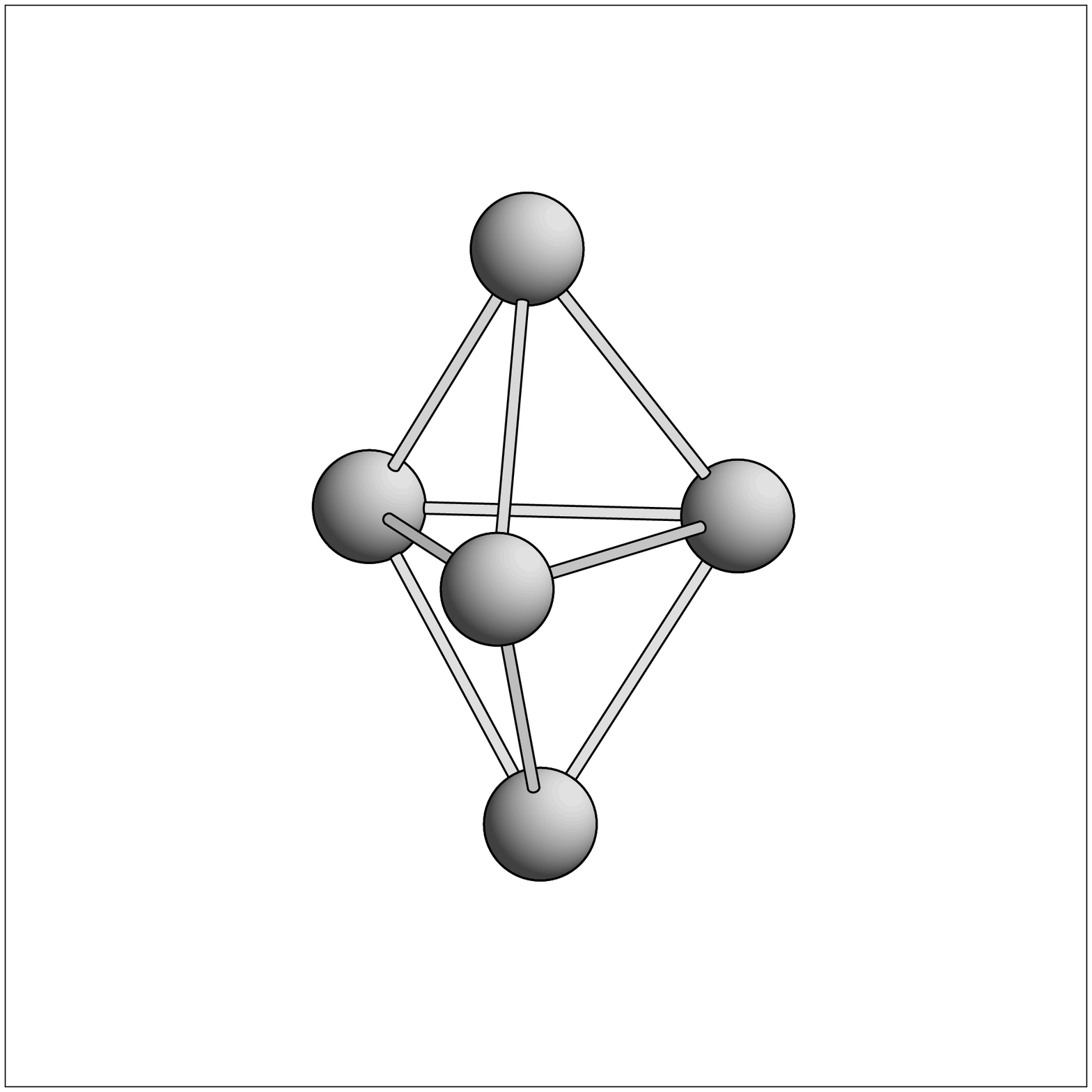,width=2.5cm}}
\put(6,0){\psfig{file=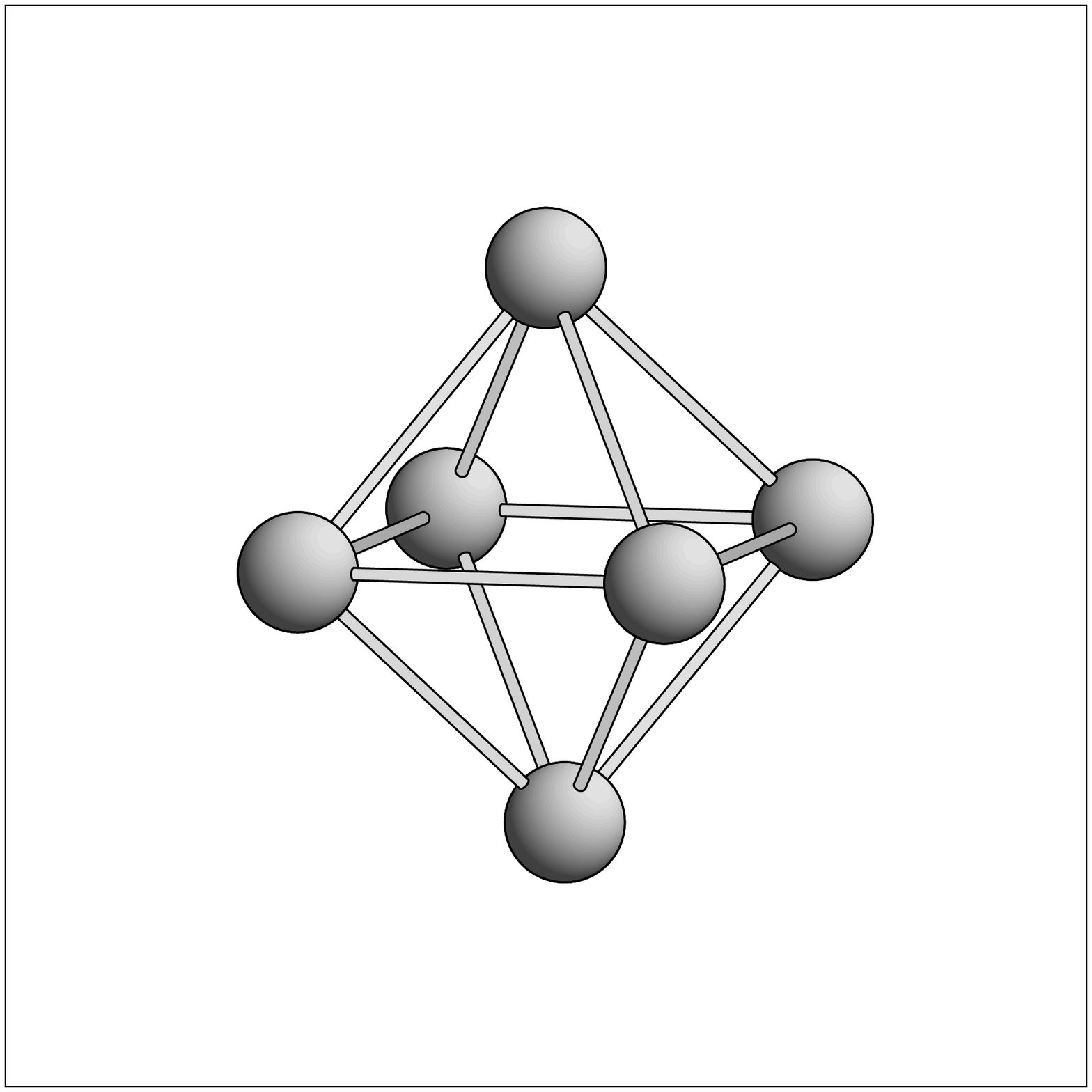,width=2.5cm}}
\put(9,0){\psfig{file=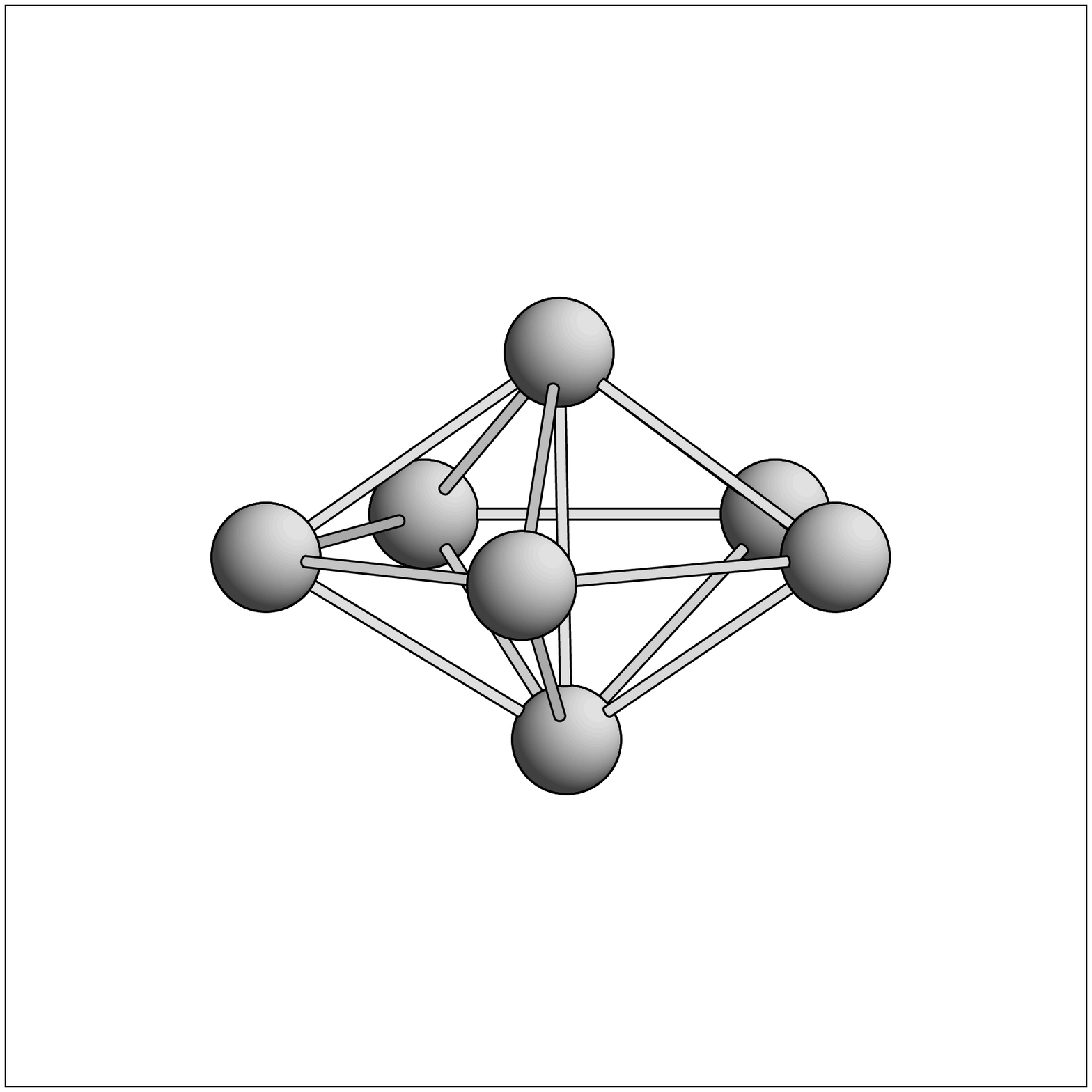,width=2.5cm}}
\put(12,0){\psfig{file=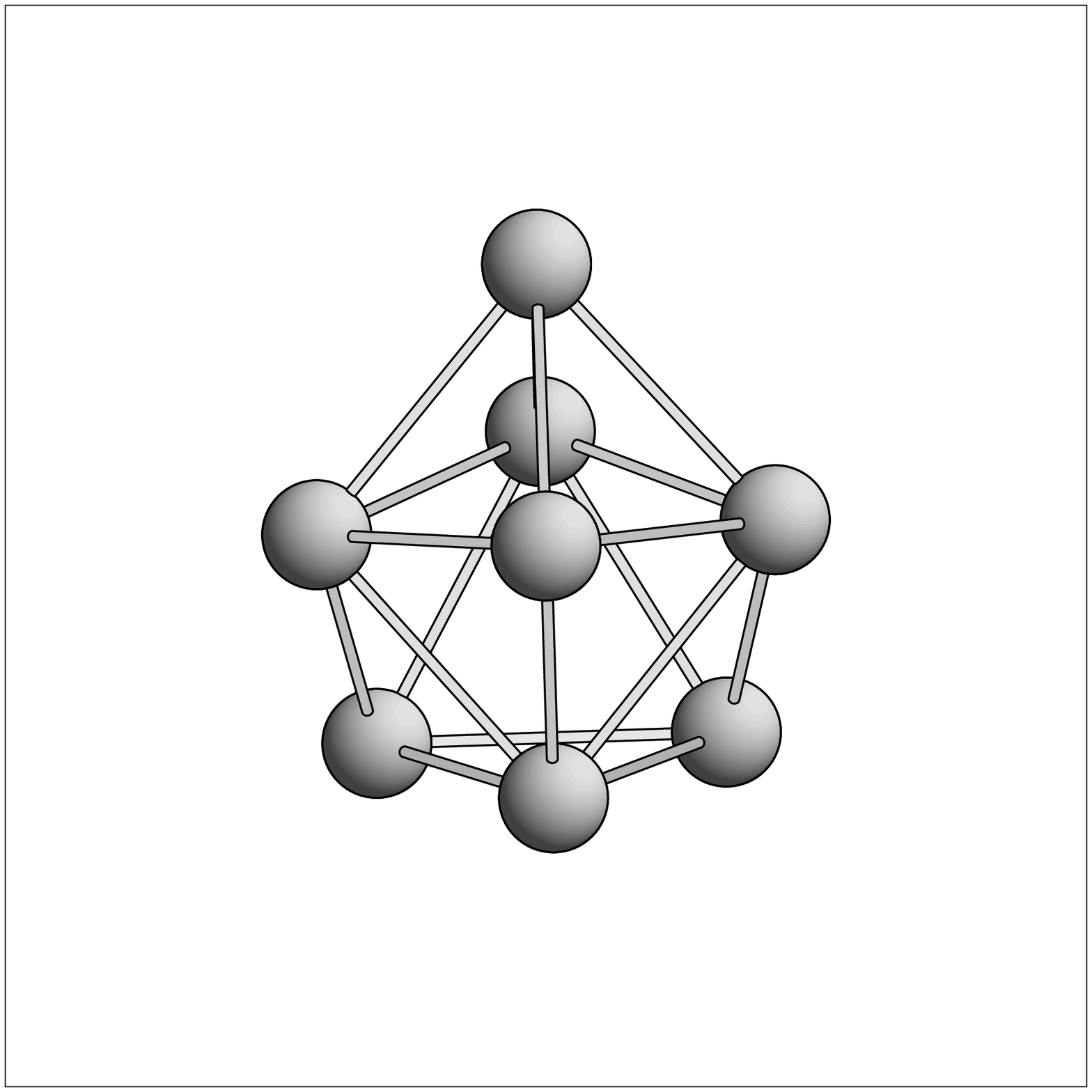,width=2.5cm}}

\end{picture}
\caption{The optimized structures for Au$_N$, $4\le N\le 8$ from (upper row) the DFT calculations,
(middle row) the DFTB calculations, and (lower row) the EAM calculations.}
\label{fig1}
\end{figure}

Next we show in Fig.\ \ref{fig1} the structures of Au$_N$ clusters with $4\le N\le8$
as obtained with the DFT calculations, the DFTB method,
and the EAM method. The results for the trimer correspond to an obtuse triangle in the DFT 
calculations, and an equilateral triangle for the EAM and the DFTB methods. 
The figure clearly illustrates the aspects we have discussed
above, i.e., the optimized structures of Au clusters result from a subtle interplay
between geometric and electronic effects. Thus, in the DFT calculations all clusters
are planar, whereas in the EAM calculations they are all three-dimensional (3D).
Moreover, it turns out that also relativistic effects are important. Including all
relativistic effects (i.e., also spin-orbit couplings) all gold clusters form planar structures 
at least up to $N$ = 13 (see Ref.\ \cite{ha03}). If the spin-orbit coupling is neglected, one obtains
3D global minima already at $N$ = 4.

\unitlength1cm
\begin{figure}[tbp]
\begin{picture}(18,8.5)
\put(2,0){\psfig{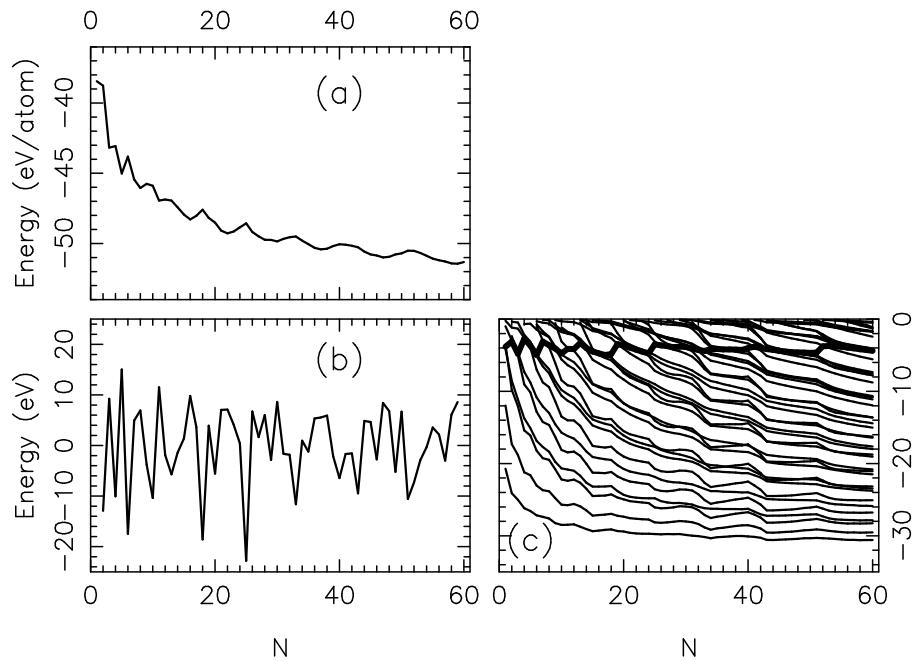}}
\end{picture}
\caption{Results from the jellium calculations on Au$_N$ clusters with $1\le N\le 60$. The
three panels show (a) total energy per atom, (b) the stability function, and
(c) the orbital energies (thin curves) and the Fermi energy (thick curve) as functions
of $N$.}
\label{fig2}
\end{figure}

The jellium model excludes packing effects and treats exclusively electronic-shell effects. 
With $r_s$ being the electron-gas parameter of the system of interest (i.e., the radius of a
sphere containing one electron) it is well-known \cite{bb72} that particularly stable clusters
(magic numbers) occur for regularly spaced spherical clusters whose radius differ by 
\begin{equation}
\Delta R = 0.603 r_s
\label{eqndrs}
\end{equation}
for not too small clusters. For even smaller clusters one finds magic numbers for clusters 
containing 2, 8, 18, 20, 34, 58, 92, 132, 138, 186, 254, 338, $\dots$ electrons (see, 
e.g., \cite{csn}). 

In Fig.\ \ref{fig2} we show results from the jellium calculations on Au
clusters, where it is assumed that each atom contributes with 11 electrons. Thus, the 
above-mentioned magic numbers are not reached for the gold clusters. Nevertheless, the
total energy per atom shows a very regular behaviour that actually can be related to 
Eq.\ (\ref{eqndrs}). By comparing the total energy with the orbital energies (also shown) 
we see that the local minima correspond to structures where the Fermi level makes a jump, i.e.,
where new electronic shells are being filled. 

In order to identify the particularly stable clusters we introduce the stability function,
\begin{equation}
\Delta_2 E(N) = E_{\rm tot}(N+1)+E_{\rm tot}(N-1)-2E_{\rm tot}(N)
\label{eqnstab}
\end{equation}
where $E_{\rm tot}(K)$ is the total energy of the Au$_K$ system. $\Delta_2 E(N)$ has local maxima
when Au$_N$ is particularly stable, i.e., when $E_{\rm tot}(N)$ is particularly low compared with
$E_{\rm tot}(N-1)$ and $E_{\rm tot}(N+1)$. This function possesses a number of maxima, as seen
in the figure, i.e., for $N=$ 3, 8, 11, 16, 21, 22, 26, 28, 30, 36, 37, 38, 47, 48, 50, 58, and 59.

\unitlength1cm
\begin{figure}[ht]
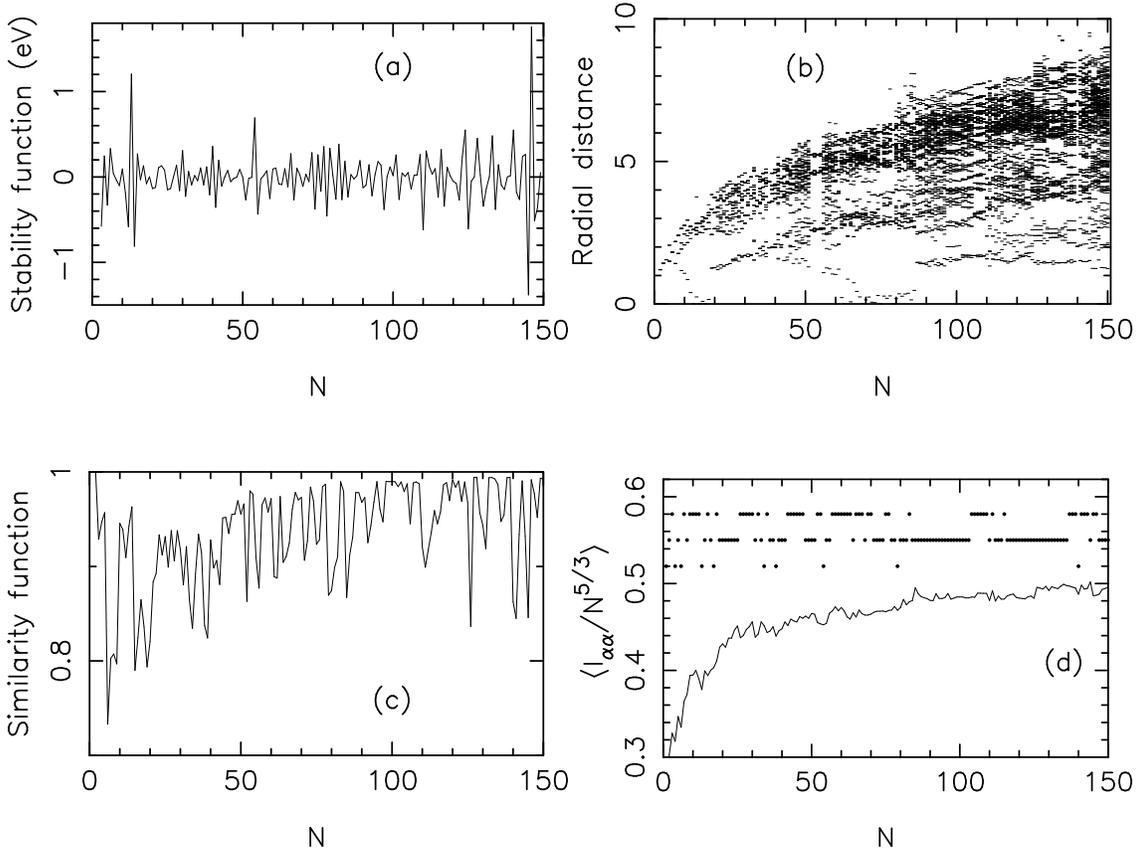

\begin{picture}(18,12.)
\put(0,6){\psfig{file=eamstab.ps,width=7.5cm}}
\put(7.5,6){\psfig{file=eamrad.ps,width=7.5cm}}
\put(0,0){\psfig{file=eamsim.ps,width=7.5cm}}
\put(7.5,0){\psfig{file=eamshap.ps,width=7.5cm}}
\end{picture}
\caption{Properties of Au$_N$ clusters from the EAM calculations. The four panels show (a)
the stability function, (b) the radial distribution of atoms, (c) the similarity function,
and (d) the shape-analysis parameters, respectively. Lengths and energies are given in \AA\, and 
eV, respectively. In (d) the upper rows show whether the clusters have an overall spherical shape
(lowest row), an overall cigar-like shape (middle row), or an overall lens-like shape (upper row).}
\label{fig3}
\end{figure}

\unitlength1cm
\begin{figure}[ht]
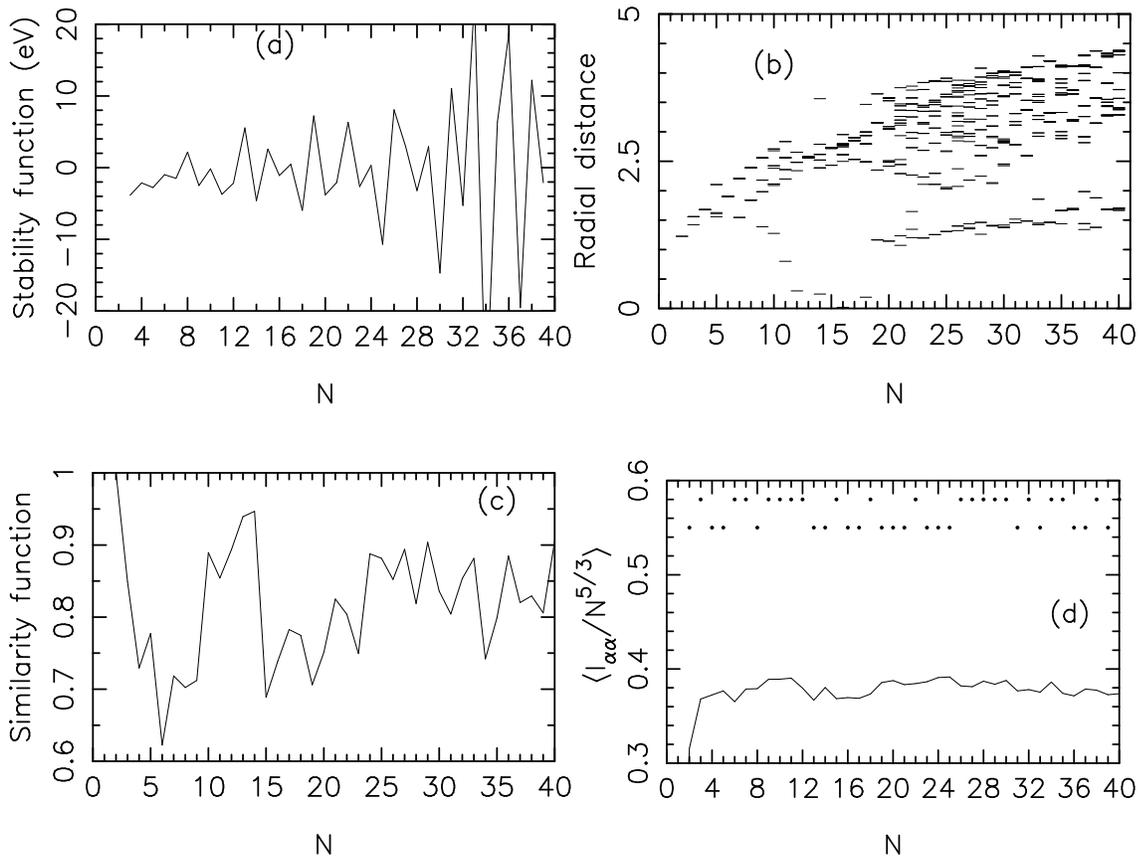

\begin{picture}(18,12.)
\put(0,6){\psfig{file=eamdftbstab.ps,width=7.5cm}}
\put(7.5,6){\psfig{file=eamdftbrad.ps,width=7.5cm}}
\put(0,0){\psfig{file=eamdftbsim.ps,width=7.5cm}}
\put(7.5,0){\psfig{file=eamdftbshap.ps,width=7.5cm}}
\end{picture}
\caption{As Fig.\ \ref{fig3}, but from the DFTB calculations with the structures of the
EAM calculations.}
\label{fig4}
\end{figure}

\unitlength1cm
\begin{figure}[ht]
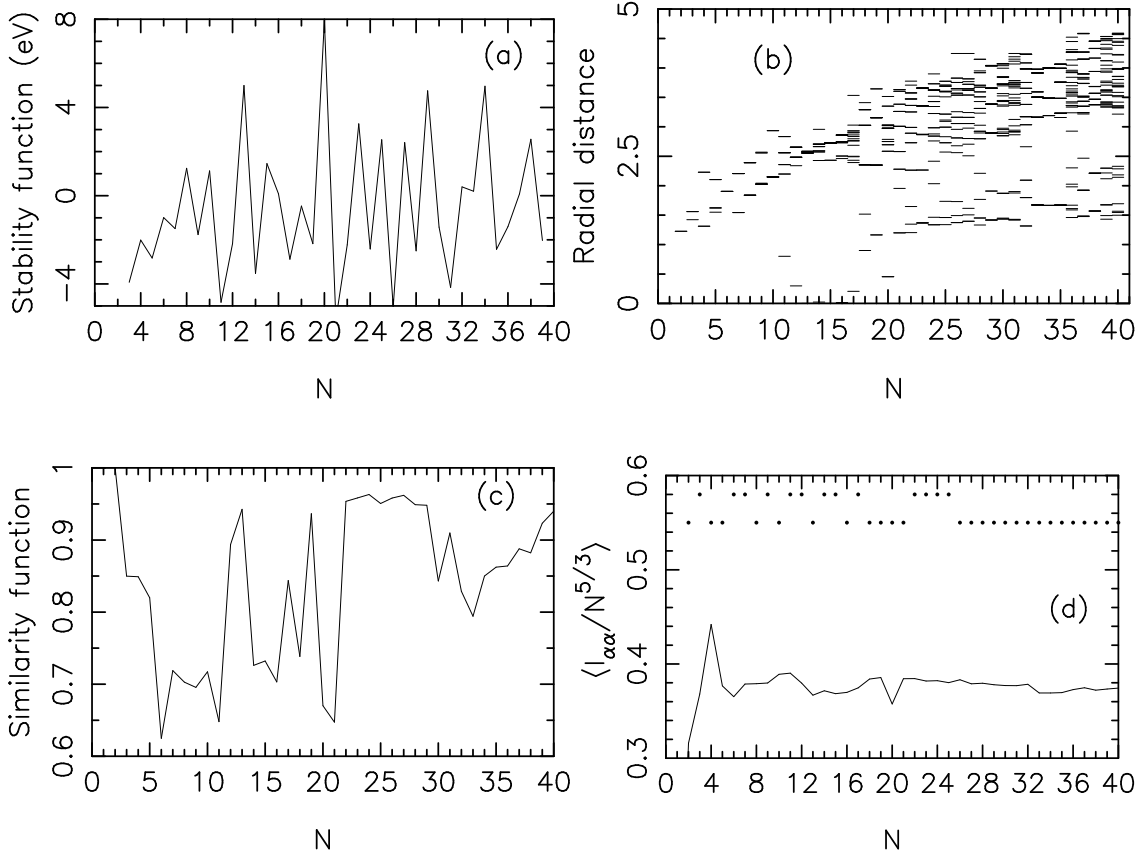

\begin{picture}(18,12.)
\put(0,6){\psfig{file=dftbstab.ps,width=7.5cm}}
\put(7.5,6){\psfig{file=dftbrad.ps,width=7.5cm}}
\put(0,0){\psfig{file=dftbsim.ps,width=7.5cm}}
\put(7.5,0){\psfig{file=dftbshap.ps,width=7.5cm}}
\end{picture}
\caption{As Fig.\ \ref{fig3}, but from the DFTB calculations with the genetic-algorithm
optimization of the structures.}
\label{fig5}
\end{figure}

We shall now compare this purely electronic description of the clusters with those obtained using
the other approaches for Au$_N$. Fig.\ \ref{fig3} shows various properties from the EAM calculations
on Au$_N$ clusters with $N$ up to 150. Here, we used our {\it Aufbau/Abbau} method in optimizing
the structure. The stability function, Fig.\ \ref{fig3}(a), has pronounced peaks at $N=13$, 54, and
146, where the structure corresponds to an icosahedron for the first two and a
decahedron for the last (actually, for $N=55$ we do not find a
icosahedron, whereas the structure of $N=54$ is the 55-atomic icosahedron without the central atom).

Further information on the structure is obtained from the radial distances of the atoms, defined
as follows. First, we define the center of the Au$_N$ cluster,
\begin{equation}
\vec R_0=\frac{1}{N}\sum_{i=1}^N \vec R_i
\label{eqnr0}
\end{equation}
and, subsequently, we define for each atom its radial distance
\begin{equation}
r_i=\vert\vec R_i-\vec R_0\vert.
\label{eqnri}
\end{equation}
In Fig.\ \ref{fig3}(b) we show the radial distances for all atoms and all cluster sizes. Each small
line shows that at least one atom for the given value of $N$ has exactly that radial distance. The 
figure shows that somewhere around $N=10$ a second shell of atoms is being built up, with a central
atom for $N=13$. Around $N=54$, there are only few values of the radial distance, i.e., the 
clusters have a high symmetry. Around $N=75$ we see that a third atomic shell is being formed.

We have earlier found \cite{aa3} that it was useful to monitor the structural development of
the isomer with the lowest total energy through the so-called similarity functions.
We shall study how clusters grow and, in particular, if the cluster with $N$ atoms can be
derived from the one with $N-1$ atoms simply by adding one extra atom. In order to
quantify this relation we consider first the structure with the
lowest total energy for the $(N-1)$-atom cluster. For this we calculate and
sort all interatomic distances, $d_i$, $i=1,2,\cdots,\frac{N(N-1)}{2}$. Subsequently we consider 
each of the $N$ fragments of the $N$-cluster that can be obtained by removing one of the atoms and
keeping the rest at their positions. For each of those we also calculate and sort all
interatomic distances $d_i'$, and calculate, subsequently, 
\begin{equation}
q=\bigg[\frac{2}{N(N-1)}\sum_{i=1}^{N(N-1)/2} (d_i-d_i')^2\bigg]^{1/2}.
\end{equation}
Among the $N$ different values of $q$ we choose the smallest one, $q_{\rm min}$,
and calculate the similarity function
\begin{equation}
S=\frac{1}{1+q_{\rm min}/u_l}
\end{equation}
($u_l$ = 1 \AA) which approaches 1 if the Au$_N$ cluster is very similar to the Au$_{N-1}$
cluster plus an extra atom. This function is shown in Fig.\ \ref{fig3}(c). We see that the 
structural development is very irregular over the whole range of $N$ that we have considered
here, with, however, some smaller intervals where $S$ is relatively large, for instance for 
$N$ slightly above 20.

Finally, we shall consider the overall shape of the clusters. As we showed in our
earlier report on Ni clusters \cite{aa3}, it is convenient to study the $3\times 3$ matrix
containing the elements
\begin{equation} 
I_{st}=\frac{1}{u_l^2}\sum_{n=1}^N (R_{n,s}-R_{0,s})(R_{n,t}-R_{0,t})
\end{equation}
with $s$ and $t$ being $x$, $y$, and $z$. The three eigenvalues of this matrix, 
$I_{\alpha\alpha}$, can be used in separating the clusters into being overall
spherical (all eigenvalues are identical), more cigar-like shaped (one eigenvalue is
large, the other two are small), or more lens-shaped (two large and one small eigenvalue).
The average of the three eigenvalues, $\langle I_{\alpha\alpha}\rangle$, is a
measure of the overall extension of the cluster. For a homogeneous sphere with $N$
atoms, the eigenvalues scale like $N^{5/3}$. Hence, we show in Fig.\ \ref{fig3}(d)
quantities related to $I_{\alpha\alpha}$ but scaled by  $N^{-5/3}$. In this figure
we also mark the overall shape of the clusters through the upper points with the lowest
row meaning spherical, the middle row meaning cigar-shaped, and the upper row meaning 
lens-shaped clusters. Some clusters with an overall spherical shape can be recognized,
that, simultaneously, are clusters of particularly high stability according to 
Fig.\ \ref{fig3}(a). 

By comparing with the results of the jellium calculations, we see that the EAM method
predicts a completely different set of particularly stable clusters. Moreover, compared
with the results of the parameter-free density-functional calculations, the EAM method
tends to produce more compact clusters. 

We shall now turn to the results obtained with the DFTB method. In Fig.\ \ref{fig4} we 
show results similar to those of Fig.\ \ref{fig3}, but obtained by letting the clusters
of the EAM calculations relax locally using the DFTB method, whereas we in Fig.\ \ref{fig5}
show the results for the DFTB calculations where the structure was optimized with the 
genetic algorithms. 

The structural information from all three sets of calculations is very similar. The radial
distances show in all cases that a second atomic shell is being constructed starting from
slightly above 10 atoms. First above $N=40$ a third atomic shell is formed [see Fig.\ 
\ref{fig3}(c)]. Moreover, the similarity function shows that clusters in a narrow window
just above $N=10$ as well as in a wider window above $N=20$ resemble each other, independently
of the theoretical approach. There are, however, some differences in the overall shape, as
seen in the shape-analysis parameters. In the EAM calculations several roughly spherical
clusters with $N\le 40$ are found, whereas the symmetry is broken when including 
electronic effects with the DFTB method, so that in these calculations no cluster is found
to have an overall spherical structure. Moreover, the structures of the DFTB calculations are
overall slightly more compact than those of the EAM calculations.

On the other hand, the stability function of Fig.\ \ref{fig4}(a) shows a much more irregular
behaviour than the one of Fig.\ \ref{fig3}(a), although the structures are very similar.
This emphasizes that electronic effects indeed are important. By comparing with Fig.\ \ref{fig5}(a)
we see that through further relaxations the stability function becomes less irregular, although
its variation (over a range of roughly 12 eV) is significantly larger than the range of the 
stability function in Fig.\ \ref{fig3}(a) (roughly 2 eV). It is interesting to observe that
in the jellium calculations (where there is per construction no structural relaxation) the
stability function spans a range of roughly 35 eV, which is comparable to the range we see
in Fig.\ \ref{fig4}(a). This means that electronic effects are very important for the stability
of the clusters, but also that through structural relaxations the role of the electronic 
effects can be somewhat reduced.

Finally, we see in Fig.\ \ref{fig1} that the DFTB calculations lead to essentially the same
(type of) structures as the EAM calculations, with $N=4$ being the only significant exception.
Here, the DFTB calculations predict a planar structure (as is the case for the relativistic
DFT calculations), whereas for larger values of $N$, non-planar structures result in all our
calculations except for the relativistic DFT calculations.

\section{Conclusions}

In this work we have discussed the interplay of electronic and packing effects in clusters.
We have used Au$_N$ clusters as a prototype in order to illustrate the effects of 
different types of approximations on the description of the interatomic interactions. We 
have, moreover, demonstrated how carefully chosen descriptors can be constructed that 
clearly grasp the essential outcomes of the calculations. Moreover, we hope also to have
demonstrated the complexities related to unbiased, accurate calculations of the properties
of not-too-small clusters.

Our results indicate the existence of a tendency for simple potentials that do not directly
include effects due to electronic orbitals (i.e., due to directed chemical bonds) to prefer
closed packed structures. Including the electronic orbitals the structures may become less
symmetric, and when increasing the accuracy of the treatment of the electronic orbitals 
and, simultaneously, the structure, the geometric arrangement of the atoms becomes less and less 
closed packed. Moreover, the electronic effects lead to a much stronger variation in stability
as a function of size. 

We have in this presentation focused on Au$_N$ clusters. These clusters are among the most
studied ones and are simultaneously very difficult to treat theoretically (as discussed in
the introduction), which was the motivation for the present work. Nevertheless, we believe
that most of our conclusions are valid also for other types of metal clusters, thus emphasizing
the importance to perform different types of calculations with different types of (approximate)
descriptions of the interatomic interactions before making finite conclusions about the 
properties of the clusters of a specific element.

\vspace{1.0cm}

\section*{Acknowledgment}

We gratefully acknowledge {\it Fonds der Chemischen Industrie} for very 
generous support. This work was supported by the SFB 277 of the University 
of Saarland and by the German Research Council (DFG) through project Sp439/14-1.


\begin{thebibliography}{99}

\bibitem{pekka} P. Pyykk\"o, Angew. Chem. Int. Ed. {\bf 43}, 4412 (2004).

\bibitem{md95} M. Dorogi, J. Gomez, R. Osifichin, R. P. Andres, and R.
Reifenberger, Phys. Rev. B {\bf 52}, 9071 (1995).

\bibitem{rlw99} R. L. Whetten, M. N. Shafigullin, J. T. Khoury, T. G. Schaaff,
I. Vezmar, M. M. Alvarez, and A. Wilkinson, Acc. Chem. Res. {\bf 32}, 397
(1999).

\bibitem{as99} A. Sanchez, S. Abbet, W. D. Schneider, H. H\"akkinen, R. N.
Barnett, and U. Landman, J. Phys. Chem. A {\bf 103}, 9573 (1999).

\bibitem{ha03} H. H\"akkinen, B. Yoon, U. Landman, X. Li, H.-J. Zhai, and L.-S.
Wang, J. Phys. Chem. A {\bf 107}, 6168 (2003).

\bibitem{sg02} S. Gilb, P. Weis, F. Furche, R. Ahlrichs, and M. M. Kappes, J.
Chem. Phys {\bf 116}, 4094 (2002). 

\bibitem{mn05} M. Neumaier, F. Weigend, and O. Hampe, J. Chem. Phys. {\bf 122},
104702 (2005).

\bibitem{br99} G. Bravo-P\'{e}rez, I. L. Garz\'{o}n, and O. Novaro, J. Mol.
Str. (Theochem) {\bf 493},
225 (1999).

\bibitem{gr00} H. Gr\"onbeck and W. Andreoni, Chem. Phys. {\bf 262}, 1 (2000).


\bibitem{ha00}  H. H\"akkinen and U. Landman, Phys. Rev. B {\bf 62}, 2287 (2000).


\bibitem{wa02}  J. Wang, G. Wang, and J. Zhao, Phys. Rev. B {\bf 66}, 035418
(2002).

\bibitem{avw05} A. V. Walker, J. Chem. Phys. {\bf 122}, 094310 (2005).

\bibitem{fr05} F. Remacle and E. S. Kryachko, J. Chem. Phys. {\bf 122}, 044304
(2005).

\bibitem{do98} J. P. K. Doye and D. J. Wales, New J. Chem. 733 (1998).

\bibitem{wi00} N. T. Wilson and R. L. Johnston, Eur. Phys. J. D {\bf 12}, 161
(2000).

\bibitem{ha97}  O. D. H\"aberlen, S. C. Chung, M. Stener, and N. R\"osch, J.
Chem. Phys. {\bf
106}, 5189 (1997).

\bibitem{clc97} C. L. Cleveland, U. Landman, T. G. Schaaff, M. N. Shafigullin,
P. W. Stephens, and 
R. L. Whetten, Phys. Rev. Lett. {\bf 79}, 1873 (1997).  

\bibitem{rb99} R. N. Barnett, C. L. Cleveland, H. H\"akkinen, W. D. Luedtke, C.
Yannouleas, 
and U. Landman, Eur. Phys. J. D {\bf 9}, 95 (1999).

\bibitem{cc97} C. L. Cleveland, U. Landman, M. N. Shafigullin, P. W. Stephens,
and 
R. L. Whetten, Z. Phys. D {\bf 40}, 503 (1997).  

\bibitem{ilg96} I. L. Garz\'on and A. Posada-Amarillas, Phys. Rev. B {\bf 54},
11796 (1996).

\bibitem{ilg98} I. L. Garz\'on, K. Michaelian, M. R. Beltr\'an, A.
Posada-Amarillas, P. Ordej\'on, E. 
Artacho, D. 
S\'anchez-Portal, and J. M. Soler, Phys. Rev. Lett. {\bf 81}, 1600 (1998).

\bibitem{ilg99} I. L. Garz\'on, K. Michaelian, M. R. Beltr\'an, A.
Posada-Amarillas, P. Ordej\'on, E. 
Artacho, D. 
S\'anchez-Portal, and J. M. Soler, Eur. Phys. J. D {\bf 9}, 211 (1999).

\bibitem{km99} K. Michaelian, N. Rend\'on, and I. L. Garz\'on, Phys. Rev. B
{\bf 60}, 2000 (1999).

\bibitem{tl00} T. X. Li, S. Y. Yin, Y. L. Ji, B. L. Wang, G. H. Wang, and J. J.
Zhao, Phys. Lett. A 
{\bf 267}, 403 
(2000).

\bibitem{sd02} S. Darby, T. V. Mortimer-Jones, R. L. Johnston, and C. Roberts,
J. Chem. Phys. {\bf 116}, 1536 (2002).

\bibitem{bf05} F. Baletto and R. Ferrando, Rev. Mod. Phys. {\bf 77}, 371 (2005).

\bibitem{nanjing} V. G. Grigoryan, D. Alamanova, and M. Springborg, Eur. Phys. J. D 
(in press).

\bibitem{gauss} M. J. Frisch, G. W. Trucks, H. B. Schlegel, G. E. Scuseria, M. A. Robb,
J. R. Cheeseman, J. A. Montgomery Jr., T. Vreven, K. N. Kudin, J. C. Burant, J. M. Millam,
S. S. Iyengar, j. Tomasi, V. Barone, B. Mennucci, M. Cossi, G. Scalmani, N. Rega, 
G. A. Petersson, H. Nakatsuji, M. Hada, M. Ehara, K. Toyota, R. Fukuda, J. Hasegawa,
M. Ishida, T. Nakajima, Y. Honda, O. Kitao, H. Nakai, M. Klene, X. Li, J. E. Knox, 
H. P. Hratchian, J. B. Cross, C. Adamo, J. Jaramillo, R. Gomberts, R. E. Strattmann,
O. Yazyev, A. J. Austin, R. Cammi, C. Pomelli, J. W. Ochterski, P. Y. Ayala, K. Morokuma,
G. A. Voth, P. Salvador, J. J. Dannenberg, V. G. Zakrzewski, S. Dapprich, A. D. Daniels,
M. C. Strain, O. Farkas, D. K. Malick, A. D. Rabuck, K. Raghavachari, J. B. Foresman, 
J. V. Ortiz, Q. Cui, A. G. Baboul, S. Clifford, J. Cioslowski, B. B. Stefanov, G. Liu,
A. Liashenko, P. Piskorz, I. Komaromi, R. L. Martin, D. J. Fox, T. Keith, M. A. Al-Laham,
C. Y. Peng, A. Nanayakkara, M. Challacombe, P. M. W. Gill, B. Johnson, W. Chen, M. W. Wong,
C. Gonzalez, and J. A. Pople, \textsc{Gaussian03}, {\it revision C.02}, Gaussian Inc.,
Wallingford, CT, 2004.

\bibitem{pbe} J. P. Perdew, K. Burke, and M. Ernzerhof, Phys. Rev. Lett. 77,
3865 (1996).

\bibitem{jpp97} J. P. Perdew, K. Burke, and M. Ernzerhof, Phys. Rev. Lett. 78,
1396 (1997).

\bibitem{jel1} W. A. de Heer, Rev. Mod. Phys. {\bf 65}, 611 (1993).

\bibitem{jel2} M. Brack, Rev. Mod. Phys. {\bf 65}, 677 (1993).

\bibitem{vo87}
A. F. Voter and S. P. Chen, in {\it Characterization of Defects in Materials},
edited by R. W. Siegal, J. R. Weertman, and R. Sinclair, MRS Symposia
Proceedings No. 82 (Materials Research Society, Pittsburgh, 1987), p. 175.

\bibitem{vo93} A. Voter, Los Alamos Unclassified Technical Report No LA-UR
93-3901 (1993).

\bibitem{vo95}
A. F. Voter, in {\it Intermetallic Compounds}, edited by J. H. Westbrook and
R. L. Fleischer (John Wiley and Sons, Ltd, 1995), Vol. 1, p. 77.

\bibitem{gs92} G. Seifert and R. Schmidt, New J. Chem. {\bf 16}, 1145 (1992).

\bibitem{gs96} G. Seifert, D. Porezag, and Th. Frauenheim, Int. J. Quant. Chem
{\bf 58}, 185 (1996).

\bibitem{aa1} V. G. Grigoryan and M. Springborg, Phys. Chem. Chem. Phys. {\bf 3}, 5125 
(2001).

\bibitem{aa2} V. G. Grigoryan and M. Springborg, Chem. Phys. Lett. {\bf 375}, 219 (2003).

\bibitem{aa3} V. G. Grigoryan and M. Springborg, Phys. Rev. B {\bf 70}, 205415 (2004).

\bibitem{ga1} J.-O. Joswig, M. Springborg, and G. Seifert, Phys. Chem. Chem. Phys. {\bf 3},
5130 (2001).

\bibitem{ga2} J.-O. Joswig and M. Springborg, Phys.~Rev. B {\bf 68}, 085408 (2003).

\bibitem{ga3} Y.~Dong and M.~Springborg, in {\it Proceedings of 3rd International 
Conference "Computational Modeling and Simulation of Materials"}, Ed. P. Vincenzini $et$ $al.$, 
Techna Group Publishers, p. 167 (2004).

\bibitem{bb72} R. Balian and C. Bloch, Ann. Phys. New York {\bf 69}, 76 (1972).

\bibitem{csn} M. Springborg, J. Phys. Cond. Matt. {\bf 11}, 1 (1999).

\end{thebibliography}
\end{document}